\documentclass[10pt,emptycopyrightspace]{ewsn-proc}

\usepackage{graphicx}
\usepackage{balance}
\usepackage{comment}
\usepackage{hyperref}
\usepackage{tikz}

\usepackage[font=bf]{caption}
\usepackage{pgfplots}
\usetikzlibrary{arrows,arrows.meta,external,colorbrewer,decorations.markings,decorations.pathmorphing,decorations.pathreplacing,external,fit,patterns,shapes,positioning}
\usepgfplotslibrary{groupplots,colormaps,colorbrewer,statistics,fillbetween}
\usepackage{pgfplotstable}
\pgfplotsset{compat=1.15}
\pgfplotsset{colormap/Set3-12}
\pgfplotsset{colormap/Spectral-5}
\pgfplotsset{cycle list/Set3-12}
\usepackage{tcolorbox}
\usepackage{xcolor}
\usepackage[T1]{fontenc}
\usepackage{pgfpages}
\usepackage{booktabs}
\usepackage{tabularx}
\usepackage{enumitem}
\usepackage{adjustbox}
\usepackage{makecell}
\usepackage{subfig}
\usepackage{multirow}
\usepackage{textcomp}
\usepackage{cite}
\usepackage{colortbl}
\usepackage{listings}
\usepackage[absolute,showboxes]{textpos}

\lstdefinestyle{codestyle} {
    backgroundcolor=\color{gray!5},
    basicstyle=\ttfamily,
    keywordstyle=\small\bfseries\color{Set3-D},
    numberstyle=\ttfamily\small\color{gray},
    numbers=left,
    numbersep=5pt,
    tabsize=2,
    showspaces=false,
    showstringspaces=false,
    otherkeywords={
        uint8_t,
        size_t,
        uint32_t,
        &
    }
}
\renewcommand{\keywords}[1]{\paragraph{Keywords} #1}
\numberofauthors{1}
\author{
\alignauthor \mbox{Lena Boeckmann\textsuperscript{1}, Peter Kietzmann\textsuperscript{1}, Leandro Lanzieri\textsuperscript{1}, Thomas C. Schmidt\textsuperscript{1}, Matthias W\"ahlisch\textsuperscript{2}}\\
  \affaddr{\textsuperscript{1}HAW Hamburg, \textsuperscript{2}Freie Universit\"at Berlin} \\
  \email{\{lena.boeckmann, peter.kietzmann, leandro.lanzieri, t.schmidt\}@haw-hamburg.de, m.waehlisch@fu-berlin.de}
}

\title{Usable Security for an IoT OS: Integrating the Zoo of Embedded Crypto Components Below a Common API}

\usepackage{pifont}

\usepackage{xspace}

\newcommand{\etal}{\textit{et al.}}
\newcommand{\eg}{\textit{e.g.,}~}

\newcommand{\one}{({\em i})\xspace}
\newcommand{\two}{({\em ii})\xspace}

\definecolor{riot-green}{RGB}{64,166,135}
\definecolor{colorDarkGray}{HTML}{333333}

\let\orgautoref\autoref
\renewcommand{\autoref}
{\def\sectionautorefname{Section}\def\subsectionautorefname{Section}\def\subsubsectionautorefname{Section}\orgautoref}

\newcommand*\circledOrange[1]{\tikz[baseline=(char.base)]{
\node[shape=circle,draw,inner sep=0pt,fill=Set3-F,text=colorDarkGray,font=\small, anchor=base,minimum width=4mm] (char) {\bfseries{\textsf{#1}}};}}

\newcommand*\circledGreen[1]{\tikz[baseline=(char.base)]{
\node[shape=circle,draw,inner sep=0pt,fill=Set3-G,text=colorDarkGray,font=\small, anchor=base,minimum width=4mm] (char) {\bfseries{\textsf{#1}}};}}

\makeatletter
\renewcommand{\paragraph}[1]{\vspace*{0.03in}\noindent{\bf #1.}\hspace{0.25ex \@plus1ex \@minus.2ex}}
\newcommand{\paragraphS}[1]{\vspace*{0.03in}\noindent{\bf #1}\hspace{0.25ex \@plus1ex \@minus.2ex}}
\makeatother

 \reversemarginpar
\begin{document}
\maketitle
\setlength{\TPHorizModule}{\paperwidth}
\setlength{\TPVertModule}{\paperheight}
\TPMargin{5pt}
\begin{textblock}{0.8}(0.1,0.05)
    \noindent
    \footnotesize
    If you cite this paper, please use the EWSN reference:
    L. Boeckmann, P. Kietzmann, L. Lanzieri, T. C. Schmidt, M. W\"ahlisch. Usable Security for an IoT OS: Integrating the Zoo of Embedded Crypto Components Below a Common API. In \emph{Proc. of EWSN}, ACM, 2022.
\end{textblock}

\begin{abstract}
IoT devices differ widely in crypto-supporting hardware, ranging from no hardware support to powerful accelerators supporting numerous of operations including protected key storage. An operating system should provide uniform access to these heterogeneous hardware features, which is a particular challenge in the resource constrained IoT.
Effective security is tied to the usability of cryptographic interfaces.
A thoughtful API design is challenging, and it is beneficial to re-use such an interface and to share the knowledge of programming embedded security widely.

In this paper, we integrate an emerging cryptographic interface into usable system-level calls for the IoT operating system RIOT, which runs on more than 240 platforms. This interface supports ID-based key handling to access key material in protected storage without exposing it to anyone. Our design foresees hardware acceleration on all available variants; our implementation integrates diverse cryptographic hardware and software backends via the uniform interface.
Our performance measurements show that the overhead of the uniform API with integrated key management is negligible  compared to the individual crypto operation. Our approach enhances the usability, portability, and flexibility of cryptographic support in the IoT.
\end{abstract}



\section{Introduction}
\label{sec:background}
\begin{figure}
    \includegraphics{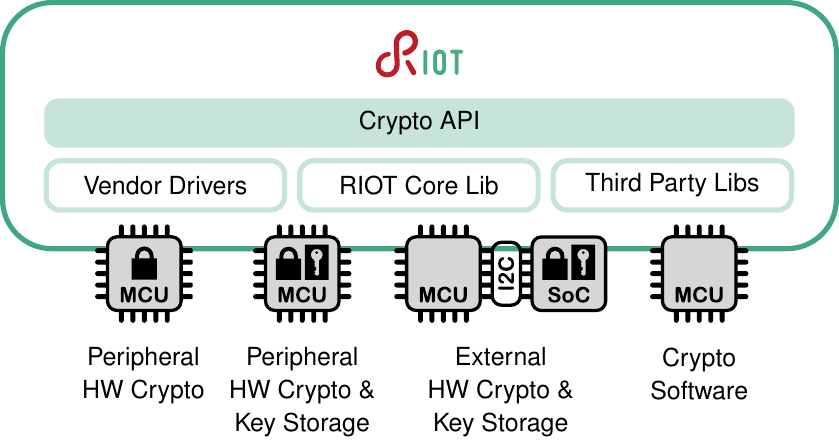}
    \caption{Variety of backends covered by the unified cryptographic API.}
    \label{fig:crypto-backends}
\end{figure}
The number of devices connected to the Internet of Things (IoT) is rapidly growing and so is its threat potential~\cite{aabbb-umb-17,kscga-atcai-19,hnkds-sunws-22}. It is an urgent demand to harden the security of the IoT ecosystem.
An important part of securing the IoT is the deployment of cryptographic operations, which often strain resources and do not comply with device constraints. Optimized libraries and crypto-supporting hardware help in mitigating resource conflicts. A proper use, however, requires specialized knowledge in selecting and programming these crypto components.

In this paper, we follow the goal of integrating an OS-level cryptographic API with interchangeable hardware and software backends in the IoT operating system RIOT.
Kietzmann \etal~\cite{kblsw-pscli-21} proposed a solution that uses the system level APIs of the cipher and hash modules that already exist in RIOT.
We argue that a common, well-known crypto API is a better choice, as it increases usability, portability, and provides interfaces for operations that are difficult to support in software on the OS level for constrained, low-end devices.

There is a need for a new architecture to support all the different hardware features optimally, while hiding its complexity at the lowest possible overhead. In our solution, the choice of crypto backends is transparent and the API abstracts specific libraries and vendor drivers away. We use semi-automatic compile time configurations to provision the available backends for each board.  This enables developers to program security without the concern about the actual support.

The remainder of this paper is structured as follows. \autoref{sec:background} discusses basic aspects of crypto-support on embedded devices together with related work and describes RIOT and its features.
\autoref{sec:api-search} discusses our requirements for a cryptographic API and explains our choice.
\autoref{sec:integration-riot} presents our implementation design and architecture.
In \autoref{sec:overhead}, we evaluate the memory and runtime overhead of our implementation. Finally, we conclude with an outlook in \autoref{sec:conclusion-outlook}.
 \section{Background and Related Work}
\subsection{Cryptographic Backends}
\label{sec:cryptographic-backends}
Different IoT devices provide various options to perform cryptographic operations as visualized in \autoref{fig:crypto-backends}. Our goal is to select  accelerating hardware backends whenever they are available and fall back to software otherwise.

\paragraph{Peripheral Hardware Accelerators}
Cryptographic co-processors implement selected operations in hardware. They are faster and consume less energy than software implementations~\cite{kblsw-pscli-21}. Some offer support for full crypto schemes, others only provide crypto primitives and require software assistance to perform compound operations. Vendors provide drivers with proprietary APIs to access hardware crypto operations. They usually accept plain text key material, which needs to be supplied by the caller.

\paragraph{Peripheral Hardware Accelerators with Key Storage}
Some platforms offer protected key storage in addition to hardware acceleration. Key material is stored in dedicated key slots, which are only accessible by the crypto processor. Driver APIs cannot input plain text keys, but identifiers or slot numbers that indirectly reference keys. The processor  then operates with the key stored in the specified slot.

\paragraph{External Crypto Devices}
External devices (\eg Secure Elements) offer tamper proof key storage and cryptographic processors to perform operations on protected key material. They connect to a chosen platform via a serial bus. Like accelerators with protected key storage they store key material in memory slots, which can only be accessed by the device processor. Different from on-chip accelerators these devices are not necessarily more efficient than software~\cite{kblsw-pscli-21}.

\paragraph{Software Implementations}
Applications running on MCUs without  cryptographic hardware features require full software implementations to execute the crypto operations.
Optimized libraries for constrained devices offer resource efficient implementations of varying degrees.
Except for Mbed TLS, which implements the ARM PSA Crypto API~\cite{arm-psacrypto-20} and accesses key material through identifier based key management, APIs of software libraries commonly accept plain text key material, which is passed to crypto functions by the caller.

\paragraph{Driver Classification}
Based on the storage of key material two classes of driver or library APIs can be identified.
\emph{Transparent Drivers} operate on plain text key material, which is passed to the implementation via input parameters.
Backends with transparent driver APIs can be easily substituted by any other transparent backend, as well as by most pure software implementations.
\emph{Opaque Drivers} interface platforms, which store their keys in protected memory. They accept identifiers or slot numbers as input, with which the crypto processor can locate the previously stored internal keys.
 To invoke an opaque driver, the location of a key must be known.
We further distinguish between opaque drivers for external devices and opaque drivers for on-chip hardware accelerators with key storage.

\subsection{RIOT}
\label{sec:riot}
RIOT~\cite{bghkl-rosos-18} is an open source operating system for low-end IoT devices that runs on architectures ranging from 8-bit to 32-bit processors.
At the time of writing, RIOT supports 240 boards, some of which offer hardware cryptography to varying extent.
A range of symmetric ciphers and hashes are available from system level software.
The \texttt{package} system can be used to add cryptographic functionality by including external libraries designed for  constrained embedded devices.
RIOT also integrates the secure element Microchip ATECC, for which a vendor driver is available as a package. Overall, RIOT provides heterogeneous backends of all classes, so that we can demonstrate integration below the cryptographic interface covering its full complexity.

For configuration, RIOT is currently adopting Kconfig~\cite{linux-kconfig-20}, a selection-based configuration system. This feature is used to select and parametrize modules and packages, which we leverage for configuring cryptographic backends.

\subsection{Related Work}
\label{sec:related-work}
\subsubsection{Crypto-Performance w/o OS Level Integration}
\label{sec:perf-no-os-level}
The performance of plain crypto-backends without OS level integration on IoT devices  has been analyzed repeatedly.
Pearson~\etal~\cite{plzdl-mhcis-19} compare peripheral- and external crypto hardware performance, deploying Espressif and Arduino libraries.
A vendor specific SDK was used by Munoz~\etal~\cite{stcdl-aruae-18} and Lachner~\etal~\cite{ld-pedpm-19} to quantify time and energy requirements of block ciphers running software and crypto-hardware. Gerez~\etal~\cite{hknld-epdat-18} use the same SDK to measure power consumption of security protocols, deploying crypto-hardware. Similarly, Mades~\etal~\cite{mejlr-tslpi-20} and Schl\"apfer~\etal~\cite{sr-sidse-19} compare the (D)TLS overhead on common IoT boards with and without secure elements and utilize yet another vendor SDK which includes Mbed TLS~\cite{arm-mbedtls-20} as a software library.
Noseda~\etal~\cite{nzsr-pasei-22} show that secure elements can increase battery life by measuring the performance of multiple SEs when performing cryptographic operations and a DTLS handshake over secure CoAP.

\subsubsection{Cryptographic Integration at OS Level}
\label{sec:os-level-integration}
Mbed OS \cite{arm-mbed-20} is an operating system with OS level crypto. It integrates the Mbed TLS \cite{arm-mbedtls-20} implementation of the cryptographic API specified by the ARM Platform Security Architecture (PSA)~\cite{arm-psa-21, arm-psacrypto-20}. Currently, hardware crypto backends can be included by providing alternative implementations of the Mbed TLS cryptographic functions.

Zephyr, FreeRTOS, and Mynewt~\cite{apache-mynewt-20} do not provide unified system level crypto interfaces. TinyCrypt~\cite{intel-tinycrypt-20} and Mbed TLS are ported to Zephyr and Mynewt. Zephyr offers a system level cipher API with access to hardware. FreeRTOS ports WolfSSL~\cite{wolfssl-21} and Mbed TLS, and implements the PKCS\#11 \cite{pkcs11-20} interface for hardware crypto support. For TLS other third-party libraries can be accessed through a TLS abstraction layer.

Security on ARM devices can be enhanced further by secure firmware (Trusted Firmware M and A (TF-M, TF-A) \cite{arm-tfm-21, arm-tfa-21}), which are also specified by PSA and allow for secure execution of cryptographic operations in isolated memory areas.
TF-M is supported by the IoT operating systems Mbed OS, Zephyr~\cite{zephyr-20} and FreeRTOS~\cite{free-rtos-20}, among others.

\subsubsection{Crypto API Design and Usability}
\label{sec:related-usability}
Green and Smith \cite{gs-dante-16} develop ten principles for cryptographic API design. They recommend  high-level operations that are easy to use without cryptographic expertise and without documentation. They should visibly handle errors, use safe defaults, and be easy to read.
Patnaik \etal \cite{phr-usadscl-19} validate Green and Smith by analysing several crypto libraries with regard to their principles.
They add that APIs should provide clear documentation containing guidelines on how to correctly perform operations and clearly mark insecure algorithms.
Mindermann \etal~\cite{mkw-huarca-18} analyze cryptographic APIs in the Rust programming language and point out the need for secure, up-to-date example code with high-level interfaces and secure defaults.
Acar \etal~\cite{abfgkms-cuca-17} study the usability and security of Python crypto APIs concluding that security and usability are inherently linked. In addition, they point out that protecting and handling key material should not be a user responsibility.
Ukrop and Matyas \cite{um-wjdcwokc-18} suggest improvements to the usability of the OpenSSL library. Whytten and Tygar \cite{wt-wjce-99} describe common problems with security in PGP 5.0 and define a usability standard.

Our subsequent process of selecting a suitable API for RIOT OS will be guided by the results  of these studies and aims at following their insights.

 \section{The Search for a Suitable Interface}
\label{sec:api-search}

\subsection{Requirements for a Cryptographic API}
\label{sec:requirements}
The objective of this work is to foster usable security for the IoT by integrating a versatile  and effective programming interface on the OS-level.
Hence, we base our requirements for such a cryptographic API on the principles and recommendations made by existing research, as well as the characteristic constraints of the IoT.

\paragraph{Hardware Platform Portability}
Applications using the interface should run on all hardware platforms supported by the OS, independent of whether specific crypto features are available in hardware.
To make this possible the interface should support implementation-agnostic development. It should abstract library and vendor specific APIs to enable easy exchange of backends that perform operations.

\paragraph{Application \& OS Portability}
Applications shall be portable to other OSes that provision the same API without changing the crypto-related code.
Thus, the API should be widely accepted and supported by other software systems.

\paragraph{Extensiveness}
A consistent cryptographic API should support all available choices of algorithms supported in hardware or software. It should allow for any combination of drivers and libraries depending on hardware capabilities.
To be compatible with platforms that offer protected key storage, the API should allow for indirect key access via identifiers as well as provide the possibility of importing and using plain text key material.

\paragraph{Usability}
The API shall guide programmers in writing secure applications, even if they are inexperienced in cryptography.
Therefore, it should have a simple, usable interface.
Calls and defaults shall be designed to avoid traps and accidental misuse.
It must be well documented and widely supported, as to facilitate the search for code examples and help.

The API should provide a way to securely handle keys and prevent their misuse. It must be possible to restrict usage of and access to keys by enforcing policies which cannot be easily overwritten.
The interface should produce meaningful and comprehensible error messages for fault resistance and easy debugging.

\subsection{The PSA Crypto API}
\label{sec:psa-crypto}
PSA Crypto is a cryptographic API specified by the \emph{ARM Platform Security Architecture (PSA) Framework} \cite{arm-psa-21, arm-psacrypto-20}. The framework provides a set of standardized resources and guidelines to facilitate the development of secure IoT systems.
It includes specifications of functional APIs for cryptography, secure storage, attestation and firmware updates. These APIs aim to be platform independent and enable developers to utilize PSA services.
The specifications are complemented by the PSA test suite \cite{arm-psatests-21}, which is open source, implemented in C language and available on GitHub. These tests can be used to verify the correct implementation of the functional APIs.
Additionally, systems developed using the PSA guidelines can be certified on multiple levels by the PSA Certified Framework \cite{arm-psacertified-21}.

The IoT OSes Mbed OS, Zephyr and FreeRTOS are already PSA certified and recently the library wolfSSL~\cite{wolfssl-21} added a wrapper to utilize PSA Crypto APIs as a backend for cryptographic operations.
These are indicators that PSA is already widely supported in the IoT industry and will be further developed and maintained in the future.

\paragraph{Usability}
PSA Crypto comes with an extensive and readable documentation, which also provides guidelines on how to correctly perform cryptographic operations and which algorithms and key types are permitted for operations. It marks insecure legacy algorithms and defines a broad range of comprehensible error values for common usage mistakes.
To prevent misuse, the specification describes numerous macros to calculate correct buffer and key sizes. It also provides single-part operations, which reduce complexity.

The API makes compromises to comply with IoT constraints and requirements.
Previous research in API usability recommends the use of high-level protocol APIs (\eg establishing an SSL/TLS session), and removing the responsibility for the choice of appropriate algorithms and parameters from the user.
PSA Crypto is a mid-level API for cryptographic operations utilized by protocols. Developers utilizing this API still need to choose secure algorithms and parameters.
This has been a concious design decision to enable applications to implement standard and custom IoT protocols on constrained devices.
It is possible to build higher level APIs on top of PSA Crypto.

\paragraph{Key Management}
The \emph{identifier} based key management takes care of storing and handling key material.
Keys can be stored either in volatile or non-volatile memory or in protected memory slots on physical devices, without exposing them to the user.
Each key is stored together with \emph{attributes}.
Those contain metadata, such as extensive usage policies to protect keys from misuse and compromisation, as well as the location of the key.
To access key material, an application needs to provide an identifier, which maps to a stored key.
The actual dispatch to a transparent or opaque driver is handled by the API implementation depending on the key location.
This way PSA supports all types of backends and drivers described in \autoref{sec:cryptographic-backends}.

\paragraph{Secure Element Handling}
Microcontrollers without hardware crypto and key storage can be extended by secure elements (SE), which provide tamper proof key storage and selected cryptographic operations.
To extend storage capacities and functionality, it may be desirable to connect multiple SEs to a platform.
From the PSA Crypto reference implementation in the Mbed TLS Library \cite{arm-mbedtls-20} we can adopt an interface for SEs which makes it possible to manage multiple external crypto devices at runtime.

\paragraph{Testing}
Part of the PSA framework is the PSA Architecture test suite, which provides extensive tests to verify the implementations of all PSA functional APIs with and without TEE integration. These tests can be integrated in RIOT using the package system.

\subsection{Potential Alternatives to PSA}
\label{sec:req-alternatives}
Competing approaches of other libraries also aim to specify generic interfaces for security services.

The \emph{PKCS\#11 Cryptographic Token Interface} (also called \emph{Cryptoki})~\cite{pkcs11-20, pkcs11-guide-14} is a platform-independent, standard programming interface for cryptographic tokens.
Tokens can be hardware security modules or software implementations and multiple tokens can be connected and accessed at runtime.
Other than PSA, PKCS\#11 is not optimized for constrained IoT systems and was not intended to be a generic cryptographic interface \cite{c-osp-03}.

The \emph{Generic Security Service API (GSS API)} \cite{RFC-2743,RFC-2744} is an IETF standard for interfacing between applications and security devices. It is mainly used by Kerberos 5 and not optimized for the IoT. \begin{figure*}
    \begin{minipage}[t]{\columnwidth}
        \includegraphics{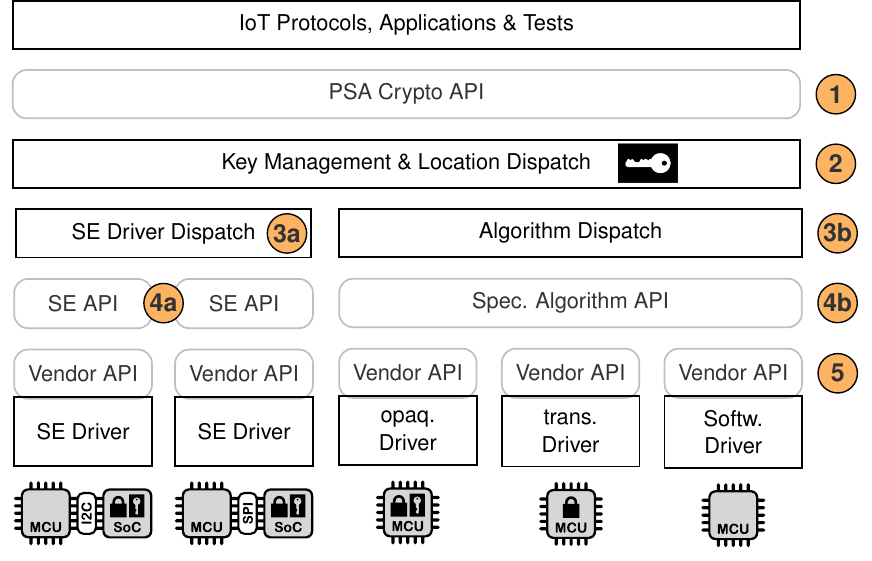}
        \caption{Layered design of cryptographic components including software and hardware implementations. The selection of features and instances is performed by a cascade of dispatchers.}
        \label{fig:psa_impl_structure}
    \end{minipage}\hspace{0.8cm}
    \begin{minipage}[t]{\columnwidth}
        \includegraphics{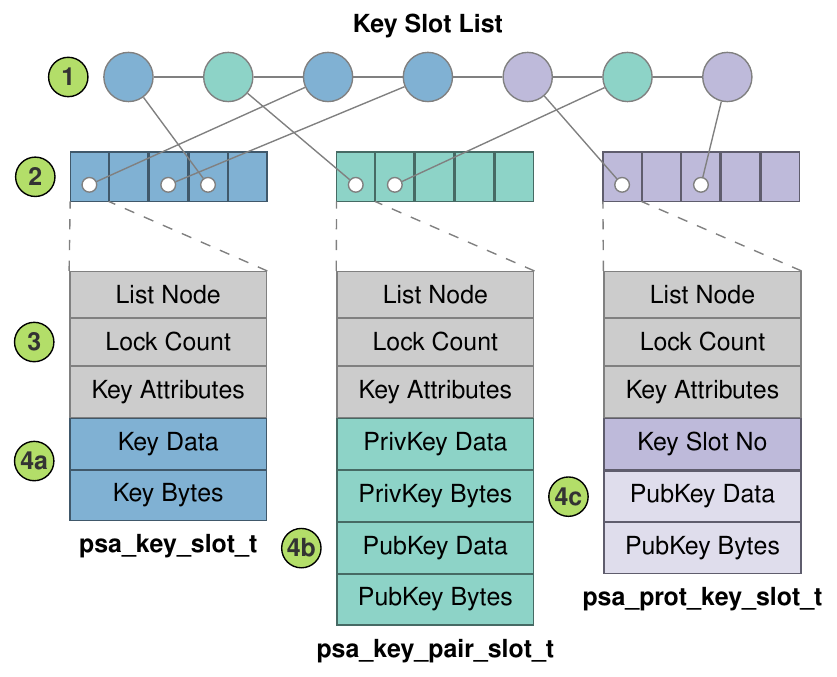}
        \caption{The three key slots types are structured differently depending on the keys they contain. They are stored in three separate arrays and abstracted by a linked key slot list.}
        \label{fig:psa_key_slot_structure}
    \end{minipage}
\end{figure*}

\section{Integrating PSA Crypto in RIOT}
\label{sec:integration-riot}
PSA Crypto enhances RIOT with a consistent system level crypto API, adding a number of features which were previously missing.
It integrates the individual software modules and libraries below a unified interface, making a full set of crypto primitives available at the OS level.
It enables transparent access to hardware accelerators and secure elements, while being highly configurable and flexible and allowing developers to tailor their application to fit the IoT constraints.
Additionally, it enhances usability and security by providing a key management module, which removes the responsibility of securely handling and storing key material from users.

\subsection{Architecture of the Crypto Subsystem}
\label{sec:implementation-arch}
The implementation of PSA Crypto in RIOT is structured in layers, which arrange as visualized in \autoref{fig:psa_impl_structure}.
The upper layer~\circledOrange{1} comprises the user facing Crypto API.
It will be directly accessed by applications for utilizing cryptographic operations.
The second layer~\circledOrange{2} consists of the \emph{key management} and \emph{location dispatcher}.
Keys are stored either in local memory or in protected hardware slots on a device.
To access keys in crypto operations, a key identifier is passed to the implementation and the key management unit will use that ID to get key metadata from memory.
The location dispatcher will check the metadata for the storage location.

If the key is stored on a secure element (SE), it will dispatch the call to the \emph{SE driver dispatcher}~\circledOrange{3a}.
SE drivers implement the generic SE interface~\circledOrange{4a} adopted from Mbed TLS.
The SE driver dispatcher will invoke the driver assigned to the key storing device, which will trigger the operation on the underlying secure element.

If the key is stored on-chip in hardware or in local memory, the location dispatcher will pass the call on to an \emph{algorithm dispatcher}~\circledOrange{3b}.
This will then check which algorithm should be performed and invoke an algorithm specific API~\circledOrange{4b}, which will be implemented by an opaque or a transparent driver.

Glue code maps the vendor-specific drivers and library APIs~\circledOrange{5} to the SE API and the algorithmic APIs.

\subsection{Key Management}
\label{sec:impl-key-management}
When using PSA Crypto, key material is handled internally underneath the API, preventing misuse and insecure storage by inexperienced users.
The key management supports functions for key creation, destruction, and export.
A key will either be created by generation, import, or copying.
Upon creation, the user specifies a set of \emph{key attributes}, which include the information about the \emph{key location} and \emph{usage policy}.
The policy determines, in which kind of operations the key can be used. It cannot be changed without destroying the key. This increases security and prevents misuse of key material.
The location encodes the actual memory location of the key. This can be either volatile or local persistent memory, or protected storage in hardware.

When generating a key, it will be stored directly in the location specified in the attributes.
If copying is permitted by the key policy, an existing key in memory can be duplicated and stored in another slot.
In both cases, the key material is not exposed to the user.
Whenever a user wants to perform operations on a key that is not stored in memory, the key must be imported first.

In all cases, an identifier is assigned to the newly created key. For volatile keys, an identifier will be generated and returned to the user.
When creating persistent keys, users can specify their own identifier.
Export functions can be used to extract existing key material from memory, if the key policy permits it, or to extract the public key of an asymmetric key pair.
A key destruction operation removes unused or compromised keys from memory.

\paragraph{Key Storage}
To store key attributes, key material and references to protected keys, we define an internal key slot data structure.
A global list contains a configurable number of key slot structures.
Key sizes can range from as small as 16 bytes for AES-128 keys up to several hundred bytes for RSA keys. For  memory efficiency, our implementation requires flexible key slot sizes.

Our solution (shown in \autoref{fig:psa_key_slot_structure}) defines three different key slot types, which are stored in three separate memory arrays~\circledGreen{2}.
The arrays are abstracted by a linked list~\circledGreen{1}.
The first three elements of all slots are equal~\circledGreen{3}: a list node is used to link the slot to the global list, the lock count keeps track of how many applications are currently reading the key and the key attributes contain the metadata of the stored key.
The structure of the actual key depends on the type of key that needs to be stored.
We differentiate between single keys, key pairs, and protected keys.

Single keys are the basic type called \texttt{psa\_key\_slot\_t}~\circledGreen{4a}. They store a single plain key and its size in bytes. They are large enough to store the largest key required by an application, \eg 16 bytes for an AES 128 key or 65 bytes for an ECC P256 public key (when importing public keys from other entities, we count them as single keys).

The type \texttt{psa\_key\_pair\_slot\_t}~\circledGreen{4b} always stores an asymmetric private and public key pair and both key sizes in bytes.

The third type \texttt{psa\_prot\_key\_slot\_t}~\circledGreen{4c} contains a reference to a key in protected memory.
When creating asymmetric key pairs on a secure element, often only the private key is stored on the device and the public key is returned by the driver.
This is why this slot type can also store a public key, if specified in the build configuration.

Using the RIOT build system configuration Kconfig (see \autoref{sec:riot-kconfig}) we ensure that key slot types are only available if needed (\eg slots for key pairs only exist when asymmetric cryptography is used).

The build configuration scheme can also specify how many key slots should be allocated for each key type depending on the application requirements.
If no key slot count is specified, the application will build with a default value.

Internally we use a linked list to handle the three key slot types, which allows for transparent access to slots.
At startup, an empty list is created for each of the three arrays, containing nodes pointing to individual array slots.
For example, when creating an asymmetric key pair, a node is removed from the empty key pair list and added to the global key slot list.
When destroying a key, the node is removed from the global list and returned to the empty list for later reuse.

\subsection{Dispatching Operations to a Backend}
\label{sec:dispatching-unit}

\paragraph{Secure Element Handling}
When utilizing secure elements, a static location value is assigned to each external device connected to a platform.
Each device driver must implement all supported methods of the generic SE interface and provide a structure with pointers to the available functions.
When booting RIOT, the OS startup function \texttt{auto\_init} initializes and registers all devices with the SE management module.
This module stores the function pointers for each device along with the location and driver context data in a global driver list.
During runtime the location value can be used to retrieve the associated driver from that list and invoke operations on the corresponding device.
This way our implementation can handle multiple SEs at the same time.

\begin{figure}[ht!]
    \includegraphics{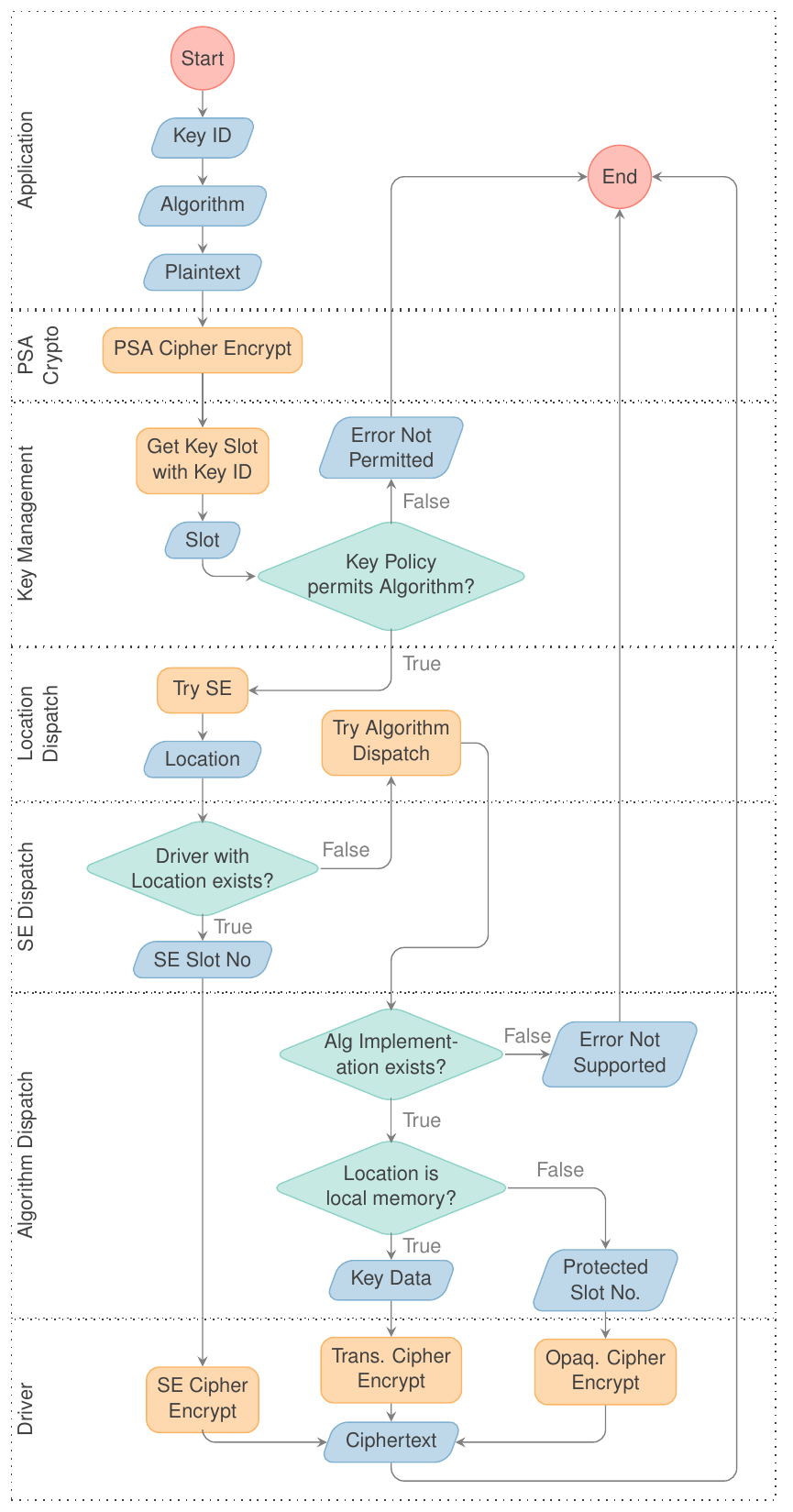}
    \caption{Flow chart of an example PSA cipher encryption using and AES 128 key in CBC mode.}
    \label{fig:psa-decision-tree}
\end{figure}

\paragraph{Dispatching Unit}
To invoke a cryptographic operation on a stored key, the user selects the desired algorithm with the key identifier, as shown in the control flow example in \autoref{fig:psa-decision-tree}.
The key management retrieves a pointer to the key slot from local memory and passes it to the location dispatcher.
The dispatcher will access the key attributes in the slot and check the location value stored in the attributes, distinguishing between secure elements and all other backends.
If the key location points to a secure element, the key slot is a protected slot and contains the key reference.

The dispatcher uses the location to retrieve the corresponding driver instance from the list of registered drivers and passes the call and the key reference to the SE.

If the location does not belong to an SE, but to some other backend, the call is passed to our algorithm dispatcher.
This dispatcher maps the key type, size, and algorithm to an algorithm-specific API (\eg \texttt{psa\_cipher\_cbc\_aes\_128\_encrypt} for an AES-128 operation in Cipher Block Chaining (CBC) mode).
If this API is implemented by a software library or driver, the contents of the key slot are passed to the corresponding backend.

Mapping the call to its algorithm-specific API allows us to combine multiple backends for different algorithms.
For example, \texttt{psa\_cipher\_cbc\_aes\_128\_encrypt} could be implemented by a hardware backend, while \texttt{psa\_cipher\_cbc\_aes\_256\_encrypt} could be implemented by a software library.

\subsection{Backend Configuration with Kconfig}
\label{sec:riot-kconfig}
The configuration of backends for PSA Crypto is done at build-time following the specific system configuration. The Kconfig facility, which has originally been developed for Linux kernel configurations, provides user-friendly access to system configurations. It can be used to select build-time options and enable and disable features via a configuration file or a graphical interface called \emph{menuconfig}.
Modules in RIOT provide Kconfig files, which define menus and configuration symbols shown by the menuconfig interface. Users can select options and features, which are then used in the build process.
Kconfig options have dependencies, which determine whether an option will be visible and selectable. Also options can be specified as default selections or be automatically selected if a certain condition is met.

We exploit these features for plugging hardware backends into PSA Crypto transparently following the capabilities of a platform.
CPUs and boards in RIOT provide their own configuration, in which they define symbols for their capabilities. If, for example, a CPU has a crypto peripheral that supports hardware cipher operations with AES keys of the size 128 bits in CBC mode, it defines the symbol \texttt{HAS\_PERIPH\_CIPHER\_AES\_128\_CBC}.

The PSA Crypto module provides its own Kconfig file, which specifies that if a symbol for hardware crypto is present at build-time, it is chosen as the default backend for the application build.
The PSA Crypto Kconfig file may also specify default software backends, which will be built in case there is no symbol for hardware crypto defined.

If developers do not wish to use the default backend for an operation, they can use menuconfig to select other available software libraries instead.
The options differ depending on the desired backend and its capabilities.
Further Kconfig can be used to optimize memory usage to the use case and the board constraints.
For example, when configuring the use of an AES 128 cipher operation, key slots will only attain the size to contain the required key.
Dedicated symbols also can be used to define the required number of key slots for each type (\eg \texttt{PSA\_ASYMMETRIC\_KEYPAIR\_COUNT=12}), and to build only the desired features.
If a user does not specify a key slot count or backend, default values are used.
 \section{Evaluation}
\label{sec:overhead}

In this section, we assess the costs and the benefits of our unified crypto systems. While costs are mainly due to memory and processing overhead from the new PSA Crypto API layer, benefits derive from reduced complexity and code duplication, but in particular from an enhanced usability of crypto operations in RIOT. We begin with quantifying the overheads.

\begin{figure}
    \begin{minipage}[t]{0.49\columnwidth}
        \includegraphics[width=\columnwidth]{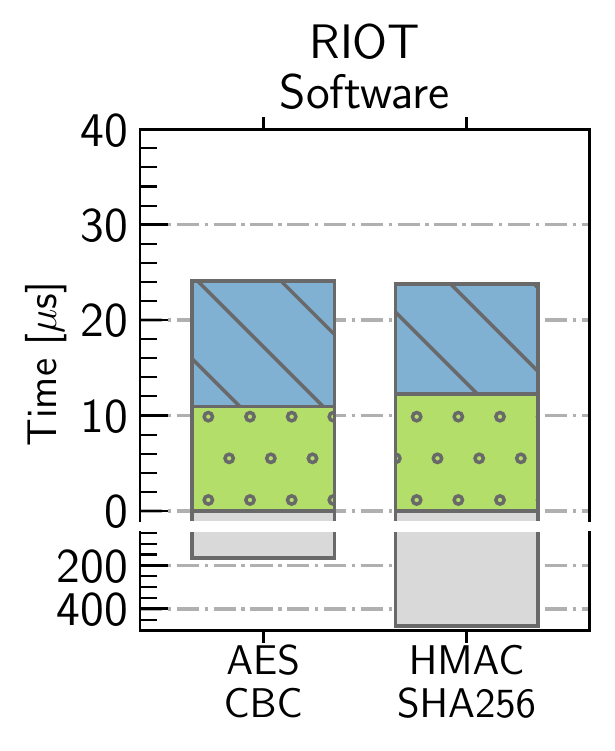}
    \end{minipage}
    \begin{minipage}[t]{0.49\columnwidth}
        \includegraphics[width=\columnwidth]{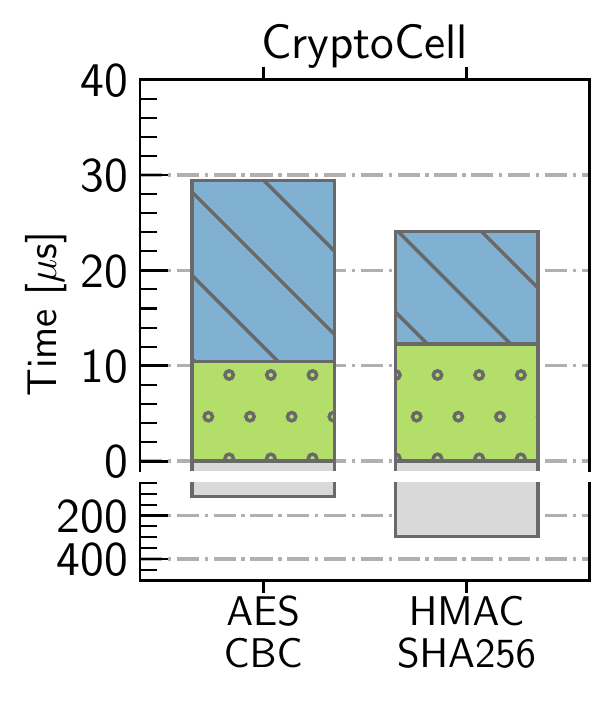}
    \end{minipage}

    \begin{minipage}[t]{0.49\columnwidth}
        \includegraphics[width=\columnwidth]{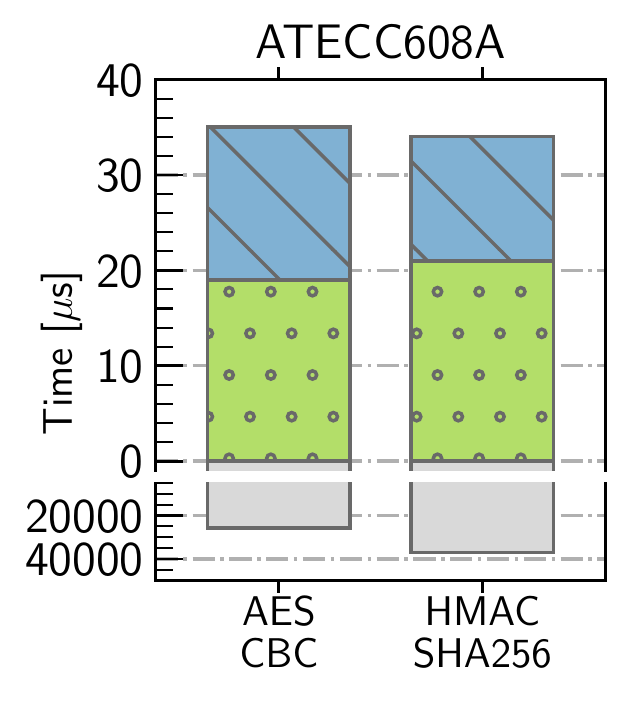}
    \end{minipage}
    \begin{minipage}[t]{0.49\columnwidth}
        \includegraphics[width=\columnwidth]{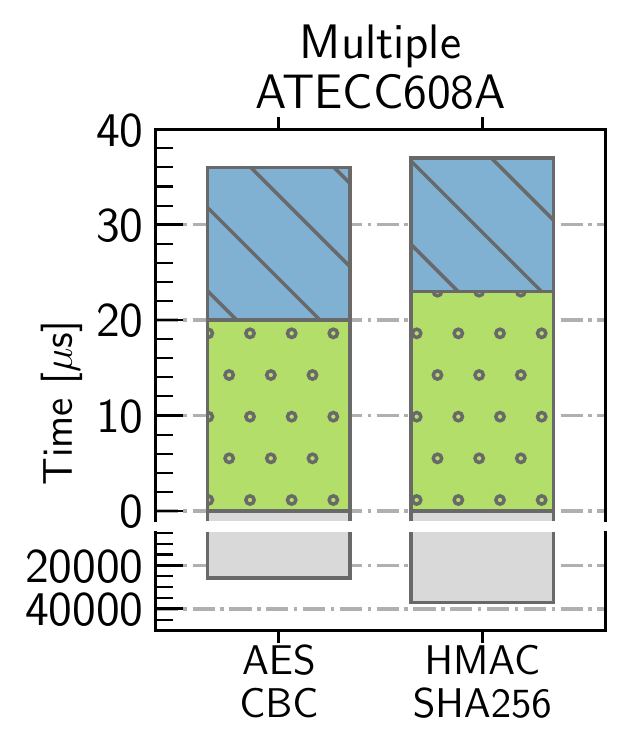}
    \end{minipage}
    \captionof{figure}{PSA Crypto API overhead compared to backend runtime for a secure element, a peripheral hardware accelerator and a software implementation of AES-128 CBC and HMAC SHA256.}
    \label{fig:symmetric-times}

    \begin{minipage}[t]{0.49\columnwidth}
        \includegraphics[width=\columnwidth]{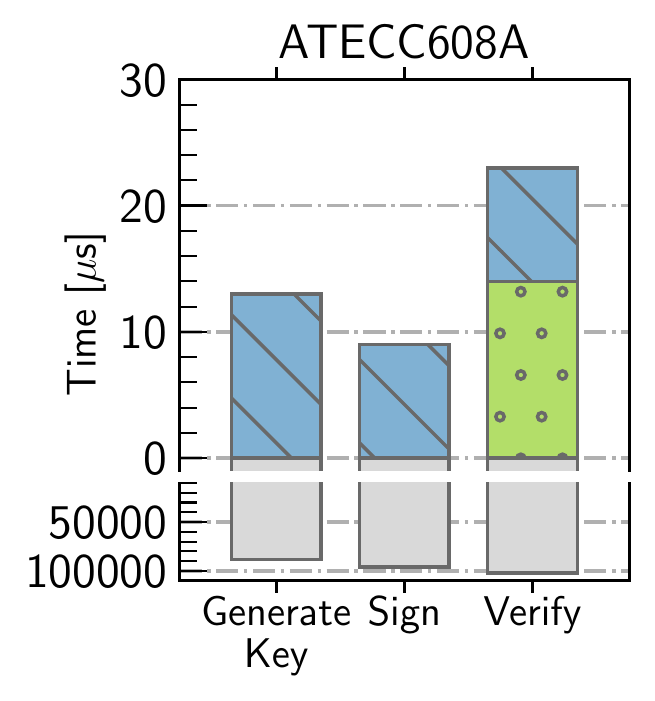}
    \end{minipage}
    \begin{minipage}[t]{0.49\columnwidth}
        \includegraphics[width=\columnwidth]{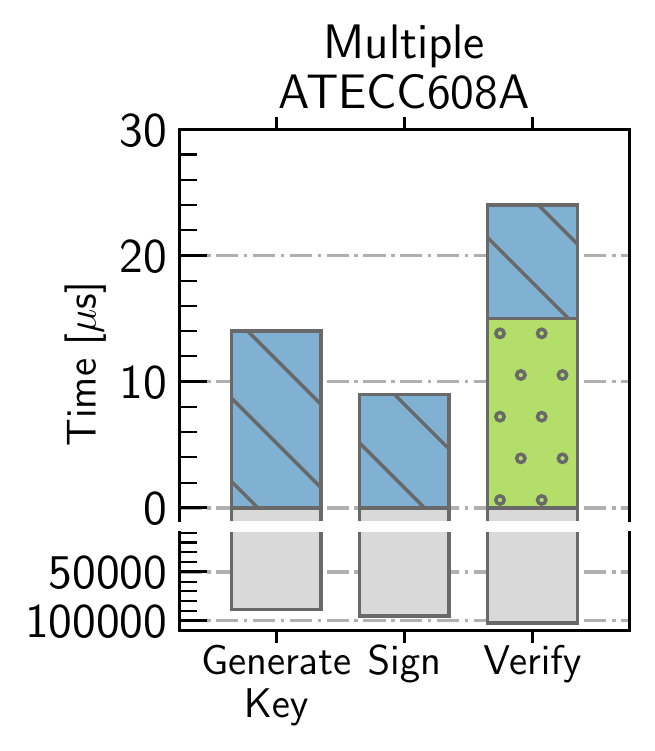}
    \end{minipage}

    \begin{minipage}[t]{0.49\columnwidth}
        \includegraphics[width=\columnwidth]{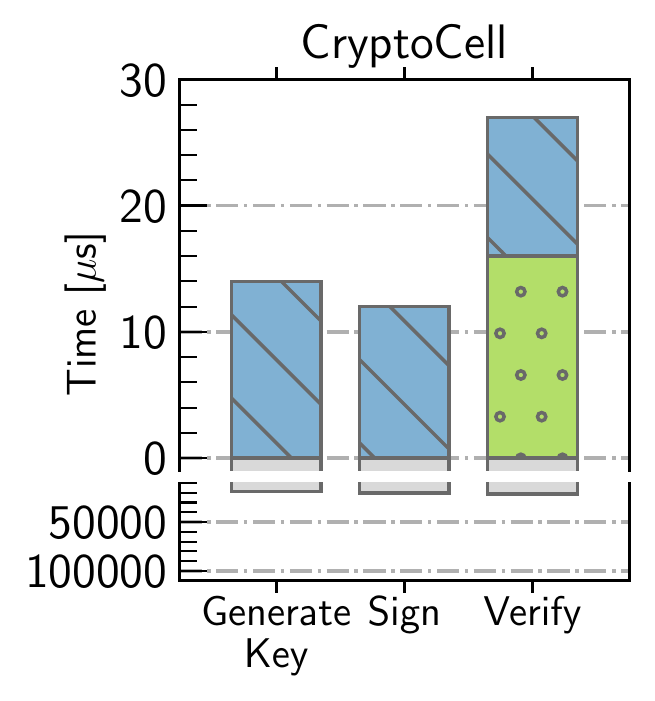}
    \end{minipage}
    \begin{minipage}[t]{0.49\columnwidth}
        \centering
        \vspace{-100pt}
        \includegraphics[trim=2pt 2pt 2pt 2pt, clip, width=0.7\columnwidth]{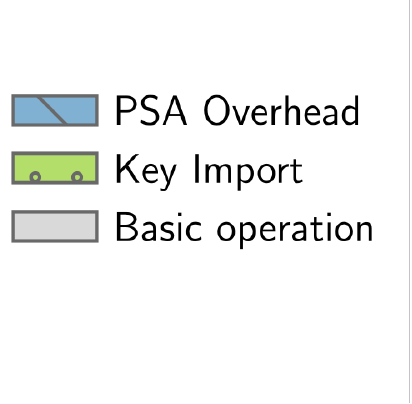}
    \end{minipage}
    \captionof{figure}{PSA Crypto API overhead compared to backend runtime for a secure element, a peripheral hardware accelerator and a software implementation of an ECDSA key generation, signature and verification.} \label{fig:asymmetric-times}
\end{figure}

\subsection{Measurement Setup}
\label{sec:eval-setup}
\begin{figure*}[t]
    \centering
        \includegraphics[width=\textwidth]{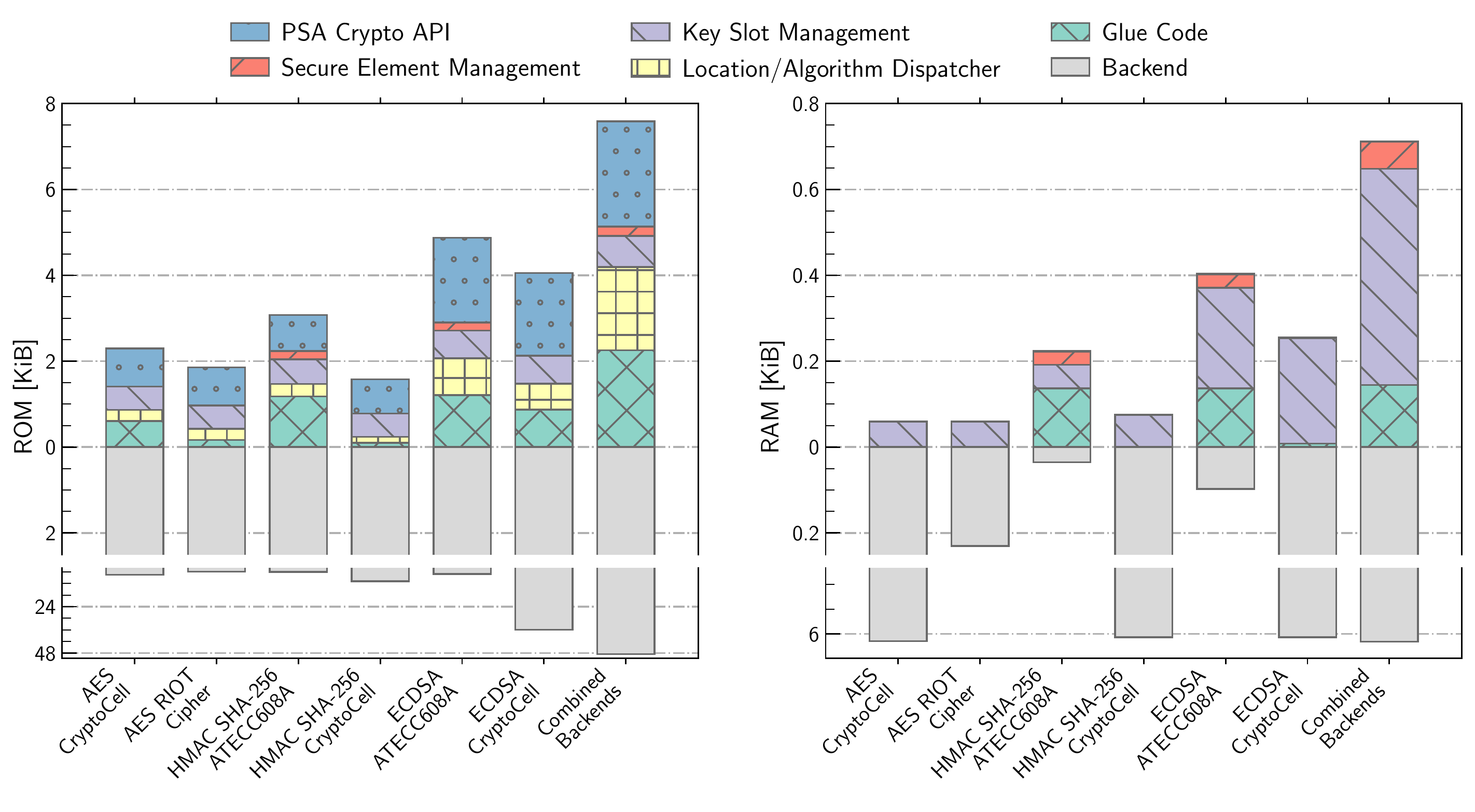}
    \caption{Memory overhead of PSA Crypto compared to various cryptographic backends in ROM and RAM.}
    \label{fig:psa-mem-overhead}
\end{figure*}

\begin{figure*}
    \begin{minipage}[t]{\columnwidth}
        \includegraphics{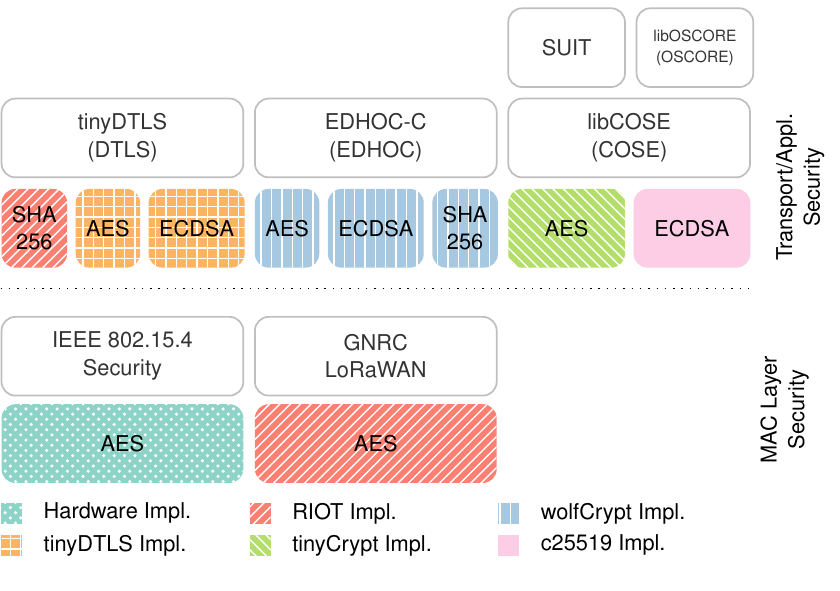}
        \caption{Example of a secure firmware configuration in RIOT when using multiple packages and modules, each of which come with their own crypto implementations.}
        \label{fig:eval-secure-stack}
    \end{minipage}
    \hspace{1cm}
    \begin{minipage}[t]{\columnwidth}
        \includegraphics{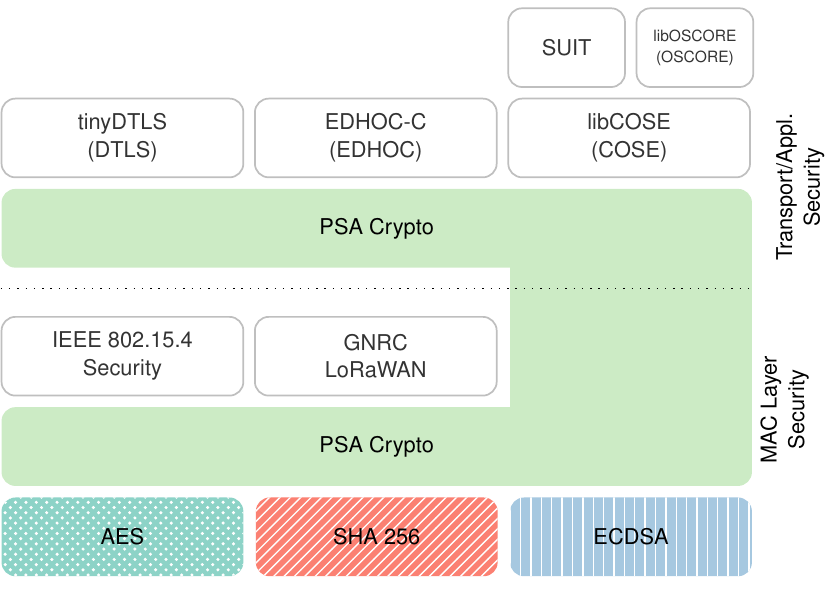}
        \caption{Reduction of code redundancy by configuring packages and modules to use the same crypto implementations through an OS level interface.}
        \label{fig:eval-secure-stack-psa}
    \end{minipage}
\end{figure*}

The cost for additional software layers on constrained embedded devices is critical.
We want to quantify these overheads for a broad range of hardware-software configurations and use cases.
We chose the Nordic nrf52840dk as our platform base, since it  provides an ARM CryptoCell 310 (CC310) peripheral accelerator and has been designed as a general purpose platform.
CryptoCell implements a large set of cryptographic operations, which can be utilized via a closed source library shipped by Nordic.
We integrate this library in RIOT and implement glue code, to map its API to our algorithm specific API, to use it as one backend for our API integration.
This approach extends transparently to other platforms.

We demonstrate the use of secure elements with PSA Crypto by connecting two external crypto-chips (Microchip ATECC608A) via I2C buses to the nrf52840dk.
The  ATECCX08A device family is commonly used in the IoT and has been widely evaluated~\cite{plzdl-mhcis-19, kblsw-pscli-21, mejlr-tslpi-20, sr-sidse-19, nzsr-pasei-22}.
The ATECC608A offers tamper proof key storage in the shape of 16 individually configurable key slots as well as a set of operations to perform on the keys.
Microchip provides the open source CryptoAuth library, which is included as a driver in PSA Crypto using the RIOT package system.
The vendor driver API is abstracted by the PSA SE interface.

We further use the HMAC and cipher operations implemented in RIOT as software backends.
Even though the nrf52840dk provides a true random number generator (TRNG), we included the RIOT random module with a deterministic SHA-256 pseudo random number generator.
This ensures constant processing time values when measuring the time overhead.

With all these components in place, our experimentation system comprises a rich variety of peripheral crypto hardware, multiple external secure elements, as well as MCU-based crypto software to assess all features of our API integration.

\subsection{Resource Metrics and Measurements}
\label{sec:eval-measured}
\paragraph{Processing Time}
We measure processing time with a logic analyzer sampling at 16 MS/s and calculate the mean value over 1000 iterations. To minimize measurement overhead we toggle I/O pins via direct register access on the nrf52840dk. To evaluate the actual overhead of PSA Crypto, we measure the complete processing time of API functions as well as the internal driver calls. We derive the PSA overhead as the difference between the external and internal measurements.
It is important to note that the symmetric operations require importing of a key prior to use. Without PSA Crypto this would only be needed when using a secure element.
We add the key import to the PSA overhead, but display it separately in Figures \ref{fig:symmetric-times} and \ref{fig:asymmetric-times}.

\paragraph{Memory Overhead}
To evaluate memory usage we analyze the compiled application ELF files and accumulate all linked objects that are associated with a crypto-implementation or are part of PSA Crypto operations.
Hereby we ignore the memory overhead of the operating system.
We analyze the sizes of individual crypto backends in ROM and RAM as well as the memory added by the API implementation.

\paragraph{Measurement Applications}
For measuring the overhead of PSA Crypto, we build applications which perform a set of symmetric and asymmetric cryptographic operations and combine multiple backends.

The \textit{HMAC SHA256 application} imports a 32 byte key to PSA and computes a MAC of a 32 byte message.
We measure the processing time of the key import and the mac computation.
As backends, we use the RIOT hash module (software), the CryptoCell peripheral and the ATECC608A.

The \textit{AES 128 CBC cipher application} imports a 16 byte key and performs the encryption on a 32 byte plaintext.
It allocates one single key slot, which can contain a 16 byte key.
We measure the processing time of the key import and the cipher encryption.
The backends used are the RIOT cipher module (software), the CryptoCell peripheral and the ATECC608A.

The \textit{ECDSA application} generates an ECC key pair using a NIST P-256 elliptic curve and uses the private key to sign a 127 byte message.
It then imports the public key to PSA and uses it to perform a verification of the signature generated before.
We measure the processing time of the key generation, the signature, the public key import and the verification procedure.
The backends used are the CryptoCell peripheral and the ATECC608A.

\subsection{Cost Results}
\label{sec:evaluation}
\paragraph{Processing Time}
\autoref{fig:symmetric-times} shows the processing time overhead of PSA Crypto in symmetric operations.
On the ATECC608A, our PSA implementation and the key import function add an overhead of less than $40\,\mu s$, which falls four orders of magnitude shorter than the actual operational processing time.

When using the RIOT software implementation, the processing time increases by $\approx 22\,\mu s$ for both the HMAC and AES operations, which is one order of magnitude less than the basic operation.

When using CryptoCell, the overhead of the HMAC operation remains constant.
For the AES operation, the overhead increases by almost $30\,\mu s$.
This is due to the initialization of the driver-specific operation context. The PSA specification requires that all contexts are initially set to zero.
The operation context used by the CryptoCell driver is approximately three times larger than the contexts used by the RIOT software implementation and the ATECC608A driver, which accounts for a longer initialization time.

The overhead added to the key import for the ATECC608A is almost twice as large as for the other backends.
To perform an AES or HMAC operation using an imported key on the ATECC608A, the key data needs import into the device TempKey register, which temporarily stores the data or intermediate  values during computations.
The implementation of the device driver requires input to TempKey of length either 32 or 64 bytes.
To ensure this, PSA copies the AES key and the HMAC key into a 32 byte buffer with padding before sending them to the device. This accounts for an increase in processing time.

When using two SEs for AES and HMAC operations, the PSA overhead increases for the key import and the actual operations, since the appropriate driver must be retrieved first.

\autoref{fig:asymmetric-times} shows the processing time of the ECDSA key generation, signing, and verifying.
The overhead of the PSA implementation and the key import remain nearly constant for both backends.
On the ATECC608A, the overhead for the key generation and the public key import are $\approx 1\,\mu s$ larger.
When using two SEs, the processing time increases by less than $1\,\mu s$ for all operations.

When using CryptoCell, the PSA overhead of the key generation, signature, and verification increases, which is due to our implementation of the algorithm dispatcher.
To dispatch calls to algorithm specific interfaces, the key type, key size and algorithm need to be mapped to the appropriate APIs, which takes $\approx 2\,\mu s$.

\paragraph{Memory Size}
\autoref{fig:psa-mem-overhead} shows the memory consumption of PSA Crypto compared to the  ROM and RAM used by drivers in the backend, omitting the memory usage of the operating system.
We built all our applications either with the CryptoCell Library, the ATECC608A driver, or the RIOT cipher module.
The rightmost bars show the memory sizes for a combination of all three backends.

The ROM used by the crypto module  depends on the backend, the amount of functions required for the operations, and the amount of glue code required to map the vendor driver APIs to the PSA APIs.
For example, when building an application for AES encryption, the PSA Crypto API, the key slot management and the dispatcher require the same amount of memory for the CryptoCell backend as well as the RIOT software implementation.
By comparison, the use of the ATECC608A almost doubles the amount of glue code and adds code for the secure element management.
Overall our PSA implementation does not exceed a size of $\approx 30\%$ of the crypto backend, while adding helpful features like the internal key management.

When combining all three backends, our implementation only requires less than 8 KB of ROM compared to over 49 KB required by the backends alone. This overhead is acceptable, comparing the provided functionality and the benefit of unifying multiple drivers and libraries with specialized vendor APIs below a consistent user interface.

In RAM, the amount of allocated memory mostly depends on the size and amount of keys used.
Operations that use small key sizes (AES and HMAC) require much less memory than asymmetric keys.
Due to our flexible key management and our individually configurable key slot sizes, we can allocate appropriate RAM efficiently at compile time.

When using secure elements, additional RAM is allocated to store SE driver instances and pointers to all available driver functions.
The application that includes a combination of all backends uses two SEs and therefore needs to store two driver instances.
This is reflected by the increase of memory consumed by the secure element management.

In summary, our cost evaluation revealed only negligible runtime overhead of our PSA Crypto integration. Memory overheads remain noticeable, but should be outweighed by \one additional functionality such as key slot management and \two code consolidation, the latter we assess in the following section.

\subsection{Code Deduplication}
\label{sec:eval-deduplication}
A secure firmware configuration in RIOT consists of multiple protocol implementations, which in many cases are provided by third party libraries.
\autoref{fig:eval-secure-stack} shows an example protocol stack built from various packages and modules, some of which bring their own implementations of cryptographic primitives.
At the transport and application layer the library tinyDTLS~\cite{eclipse-tinydtls-22} provides an implementation for Datagram Transport Layer Security (DTLS) and brings its own AES and ECDSA modules.
LibCOSE~\cite{libcose-21} implements the CBOR Object Signing and Encryption (COSE) standard optimized for constrained devices and can use different crypto backends for AES and ECDSA operations, (\eg tinyCrypt~\cite{intel-tinycrypt-20, c25519-14}).
EDHOC-C~\cite{openwsn-edhocc-22} provides a lightweight Diffie-Hellman key exchange over COSE and supports different crypto backends. In this example the package is built with wolfSSL, utilizing the AES, ECDSA and SHA-256 implementations.

At the MAC layer the IEEE 802.15.4 security module is able to use hardware acceleration for some radio modules. When using radios without access to hardware crypto, it falls back to the RIOT cipher software implementation, which is also used by the GNRC LoRaWAN implementation.

None of those modules and packages provides key management and storage functions, the implementation of which are pushed to application developers.

As apparent from \autoref{fig:eval-secure-stack}, deploying all the modules together results in five different AES implementations, two hash implementations, and three ECDSA implementations, with a combined $\sim 50$~kB of ROM usage (see \autoref{tab:secure-stack}).
Additional mechanisms for key management are also needed.
Multiple implementations require higher maintenance effort and may be prone to errors.
In the IoT, memory is a constrained resource, making the reduction of code redundancy a primary goal.

\autoref{fig:eval-secure-stack-psa} shows how the different crypto backends can be consolidated by providing a consistent OS level crypto interface that only uses one backend per algorithm.
Thanks to the internal key storage, additional key management becomes obsolete.
By modifying the packages and modules to use PSA Crypto instead of their own crypto implementations or additional libraries, code duplication can be largely avoided.
We can reduce the $\sim50$~kB of ROM usage to $\sim7$~kB by employing only the hardware accelerator, the RIOT Hash implementation and the wolfCrypt ECDSA implementation.
As  already shown, PSA adds less than 8 kB of ROM when combining multiple backends, while providing additional features like key management.
In this firmware example we demonstrate that using PSA Crypto with only one backend per operation can reduce ROM usage by $70 \%$.

\begin{table}
    \centering
    \resizebox{\columnwidth}{!}{
    \begin{tabular}{llr} \toprule
        Backend & Modules & ROM Usage $[kB]$ \\
        \midrule
        \rowcolor{gray!30}tinyDTLS & RIOT SHA-256, own ECDSA & $14$\\
        EDHOC-C & wolfCrypt (AES, ECDSA, SHA-256) & $22$ \\
        \rowcolor{gray!30}libCOSE & tinyCrypt AES, c25519 ECDSA & $8$\\
        \makecell[l]{IEEE 802.15.4\\Security} & AES Hardware Accelerator & $1$\\
        \rowcolor{gray!30}GNRC & RIOT Cipher AES & $5$\\
        \midrule
        Sum & & $50$\\
	\midrule \midrule
    \rowcolor{gray!30}Deduplicated Backends & \makecell[l]{AES HW Accelerator,\\wolfCrypt ECDSA, RIOT SHA-256} & $7$ \\
	PSA Crypto & integrated & $8$\\\midrule
    Sum & & $15$\\
        \bottomrule
    \end{tabular}
    }
	\caption{Combined ROM usage of all crypto modules of our sample configuration (cf. \autoref{fig:eval-secure-stack}) compared with the PSA Crypto integration.}
    \label{tab:secure-stack}
\end{table}

\subsection{Usability}
\begin{figure}
\begin{lstlisting}[language=C]
extern uint8_t * key;
extern size_t key_size;

int status;
uint8_t plaintext[] = {
            0x00, 0x01, 0x02, 0x03,
            0x04, 0x05, 0x06, 0x07,
            0x08, 0x09, 0x0A, 0x0B,
            0x0C, 0x0D, 0x0E, 0x0F };
uint8_t iv[16];
uint8_t output[32];
size_t output_length;
size_t size;
size_t offset = 0;
size_t length = sizeof(plaintext);

SaSiAesUserContext_t ctx;
SaSiAesUserKeyData_t user_key;
user_key.pKey = key;
user_key.keySize = key_size;

random_bytes(iv, sizeof(iv));
status = SaSi_AesInit(
            &ctx,
            SASI_AES_ENCRYPT,
            SASI_AES_MODE_CBC,
            SASI_AES_PADDING_NONE);
status = SaSi_AesSetKey(
            &ctx,
            SASI_AES_USER_KEY,
            &user_key,
            sizeof(user_key));
status = SaSi_AesSetIv(&ctx, iv);

do {
    if (length > MAX_AES_BLOCK) {
        size = MAX_AES_BLOCK;
        length -= MAX_AES_BLOCK;
    }
    else {
        size = length;
        length = 0;
    }
    status = SaSi_AesBlock(
                &ctx,
                (plaintext + offset),
                size,
                (output + offset));
    offset += size;
} while ((length > 0));

status = SaSi_AesFinish(
            &ctx, length,
            plaintext,
            sizeof(plaintext),
            output,
            &output_length);
\end{lstlisting}
    \caption{Example for an AES-128 CBC operation using the CryptoCell driver directly. The key is stored externally in RAM.}
    \label{fig:cryptocell-aes}
\end{figure}

\begin{figure}[ht!]
\begin{lstlisting}[language=C]
extern psa_key_id_t id;

psa_status_t status;
psa_algorithm_t algorithm =
            PSA_ALG_CBC_NO_PADDING;

uint8_t plaintext[] = {
            0x00, 0x01, 0x02, 0x03,
            0x04, 0x05, 0x06, 0x07,
            0x08, 0x09, 0x0A, 0x0B,
            0x0C, 0x0D, 0x0E, 0x0F };

size_t output_size =
        PSA_CIPHER_ENCRYPT_OUTPUT_SIZE(
            PSA_KEY_TYPE_AES,
            PSA_ALG_CBC_NO_PADDING,
            sizeof(plaintext))
uint8_t cipher_out[output_size];
size_t output_len;

status = psa_cipher_encrypt(
            id, algorithm,
            plaintext,
            sizeof(plaintext),
            cipher_out,
            output_size,
            &output_len);
\end{lstlisting}
    \caption{Example for an AES-128 CBC operation using PSA. The key has been previously imported and can be accessed by an ID.}
    \label{fig:psa-aes}
\end{figure}

We finally want to discuss how the integration of PSA Crypto in RIOT decreases code complexity and thereby enhances usability for security.
\autoref{fig:cryptocell-aes} displays a code example which is required to perform an AES-128 CBC operation on a previously generated key, when using the driver of the CryptoCell hardware accelerator directly.
First, the developer needs to take care of key storage and to provide access to the key material (line 1).
CryptoCell does not determine, whether this particular key is permitted for this operation, so the user has to keep track of the permissions.
She needs to know that a randomly generated initialization vector (IV) is required for this operation along with its required size.
She must generate the IV using some random number generator (RNG), such as the RIOT random module (line 22) \cite{ksw-gpngi-21}.
Now she needs to initialize an operation context (line 23), set the key (line 28), and set the previously generated IV (line 33).
She can then perform the cipher operation on the plaintext. To do this, she must keep track of how many bytes the AES block function (line 35 - 50) can compute, in case she needs to perform the operation more than once.
Finally, she can finalize the operation (line 52) to obtain the cipher text.

When performing a cipher operation using PSA Crypto with the CryptoCell backend (\autoref{fig:psa-aes}), the driver specific functionality is hidden from the developer.
Also, critical requirements like storing the key and checking its permissions, as well as generating and setting an IV are handled internally.
The extensive documentation describes the correct flows for operations and guides developers in writing their applications.
This way misuse and errors can be prevented.

The user needs to provide the desired algorithm and the identifier of a previously generated or imported key (lines 1 and 4).
The key is stored within the PSA implementation along with its type, size and usage policy, which determines the algorithms of its use.
If there is no permission to use this key in a symmetric encryption in CBC mode without padding, an error will be returned.
A macro defined by the PSA specification can be used to determine the correct size of the output buffer (line 13), reducing errors on the developer side.
The user can then perform the encryption (line 21), which internally retrieves the key material, checks the usage policy and generates a randomized IV with a previously configured RNG.
Due to our configuration system, suited implementations are chosen for the RNG, \eg hardware implementations, if available.
When building the application, the user must specify, what algorithms should be supported by the build, resulting in a binary that only includes the needed functionality and nothing more.

When the user wants to switch to another backend, she does not have to familiarize herself with another vendor API or change her code. She can simply rebuild the application with a different configuration.
 \section{Conclusion and Outlook}
\label{sec:conclusion-outlook}

Security must be usable for embedded application programmers to ensure common standards throughout the IoT. For  heterogeneous and constrained embedded systems, security can be enhanced by hardware accelerated cryptography, optimized libraries, and secure key storage.
An IoT operating system should support the use of all available implementations while increasing usability for developers. This is of particular importance for increasingly popular solutions on the application layer such as content object security~\cite{gasw-cosit-22}.

In this work, we motivated the use of PSA Crypto as a system level API, because it established as a versatile and easy-to-use interface for the IoT.
We integrated PSA Crypto to enhance the cryptographic services of RIOT with features that have been previously missing at the OS level, and to add an identifier based key management, which allows for the integration of cryptographic backends with and without protected key storage.
We demonstrated that our architecture and implementation of PSA Crypto provides transparent access to hardware and software backends with low overhead in processing time and memory.
Furthermore, we showed that the new API simplifies the use of cryptographic operations for the user.
Targeting a typical RIOT firmware configuration, we identified that the use of PSA Crypto can reduce ROM overhead by $70 \%$.

Our current implementation supports key storage in volatile memory, only.
In future work, we will extend it with secure key storage in persistent memory.
Additionally, we plan on integrating trusted execution environments (TEE) in RIOT as possible backends for PSA Crypto.
This will enable an easy support for hardware-based memory isolation and code execution in protected environments. It will further enhance the security features of RIOT OS.

For reproducibility our code is available on \url{https://github.com/inetrg/EWSN-PSA-CRYPTO-22}.
The implementation will be gradually integrated into RIOT.
 \section*{Acknowledgements}\label{sec:acknowledgements}
We would like to thank our anonymous reviewers for their valuable feedback.
Also we thank the developers at ARM and mbedTLS, who kindly answered our questions about the API design and their reference implementation.
This work was partly supported by the German Federal Ministry of Education and Research \href{https://www.bmbf.de/}{BMBF} within the project PIVOT.
 \newpage
\balance
\bibliographystyle{abbrv}
\bibliography{own,rfcs,ids,ngi,iot,layer2,meta,complexity,internet,security,programming}

\begin{thebibliography}{10}

\bibitem{abfgkms-cuca-17}
Y.~Acar, M.~Backes, S.~Fahl, S.~Garfinkel, D.~Kim, M.~L. Mazurek, and
  C.~Stransky.
\newblock {Comparing the Usability of Cryptographic APIs}.
\newblock In {\em Proc. of the IEEE Symposium on Security and Privacy (SP
  '17)}, pages 154--171, Los Alamitos, CA, USA, 2017. IEEE Computer Society.

\bibitem{free-rtos-20}
{Amazon Web Services}.
\newblock {FreeRTOS Real-time operating system for microcontrollers}.
\newblock \url{https://www.freertos.org/}, last accessed 30-11-2020, 2020.

\bibitem{aabbb-umb-17}
M.~Antonakakis \emph{et~al.}.
\newblock {Understanding the Mirai Botnet}.
\newblock In {\em 26th {USENIX} Security Symposium ({USENIX} Security 17)},
  pages 1093--1110, Vancouver, BC, Aug. 2017. {USENIX} Association.

\bibitem{apache-mynewt-20}
{Apache Software Foundation}.
\newblock {Apache Mynewt}.
\newblock \url{https://mynewt.apache.org}, last accessed 07-17-2020, 2020.

\bibitem{arm-mbed-20}
{ARM Ltd.}
\newblock {Mbed OS}.
\newblock \url{https://www.mbed.com}, last accessed 07-17-2020, 2020.

\bibitem{arm-mbedtls-20}
{ARM Ltd.}
\newblock {Mbed TLS}.
\newblock \url{https://tls.mbed.org}, last accessed 07-17-2020, 2020.

\bibitem{arm-psacrypto-20}
{ARM Ltd.}
\newblock {PSA Cryptography API 1.0}.
\newblock \url{https://armmbed.github.io/mbed-crypto/html/index.html}, last
  accessed 09-28-2021, 2020.

\bibitem{arm-psa-21}
{ARM Ltd.}
\newblock {ARM Platform Security Architecture}.
\newblock
  \url{https://developer.arm.com/architectures/architecture-security-features/platform-security},
  last accessed 09-28-2021, 2021.

\bibitem{arm-psacertified-21}
{ARM Ltd.}
\newblock {ARM PSA Certified}.
\newblock \url{https://www.psacertified.org}, last accessed 09-28-2021, 2021.

\bibitem{arm-tfa-21}
{ARM Ltd.}
\newblock {ARM Trusted Firmware A}.
\newblock \url{https://trustedfirmware-a.readthedocs.io/en/latest/}, last
  accessed 10-07-2021, 2021.

\bibitem{arm-tfm-21}
{ARM Ltd.}
\newblock {ARM Trusted Firmware M}.
\newblock \url{https://tf-m-user-guide.trustedfirmware.org}, last accessed
  09-28-2021, 2021.

\bibitem{arm-psatests-21}
{ARM Ltd.}
\newblock {PSA Functional APIs Architecture Test Suite}.
\newblock
  \url{https://github.com/ARM-software/psa-arch-tests/tree/master/api-tests/dev_apis},
  last accessed 10-07-2021, 2021.

\bibitem{bghkl-rosos-18}
E.~Baccelli, C.~G{\"u}ndogan, O.~Hahm, P.~Kietzmann, M.~Lenders, H.~Petersen,
  K.~Schleiser, T.~C. Schmidt, and M.~W{\"a}hlisch.
\newblock {RIOT: an Open Source Operating System for Low-end Embedded Devices
  in the IoT}.
\newblock {\em IEEE Internet of Things Journal}, 5(6):4428--4440, December
  2018.

\bibitem{c-osp-03}
J.~Clulow.
\newblock {On the Security of PKCS {\#}11}.
\newblock In {\em Cryptographic Hardware and Embedded Systems (CHES '03)},
  pages 411--425, Berlin, Heidelberg, 2003. Springer-Verlag.

\bibitem{c25519-14}
{Daniel Beer}.
\newblock {Curve25519 and Ed25519 for low-memory systems}.
\newblock \url{https://www.dlbeer.co.nz/oss/c25519.html}, last accessed
  07-28-2022, 2014.

\bibitem{eclipse-tinydtls-22}
{Eclipse Foundation}.
\newblock {Eclipse tinyDTLS}.
\newblock \url{https://github.com/eclipse/tinydtls}, last accessed 05-10-2022,
  2017.

\bibitem{hknld-epdat-18}
A.~H. Gerez, K.~Kamaraj, R.~Nofal, Y.~Liu, and B.~Dezfouli.
\newblock {Energy and Processing Demand Analysis of TLS Protocol in Internet of
  Things Applications}.
\newblock In {\em International Workshop on Signal Processing Systems (SiPS
  '18)}, pages 312--317, NJ, USA, 2018. IEEE.

\bibitem{gs-dante-16}
M.~Green and M.~Smith.
\newblock {Developers are Not the Enemy!: The Need for Usable Security APIs}.
\newblock {\em IEEE Security and Privacy}, 14(5):40--46, 2016.

\bibitem{gasw-cosit-22}
C.~G{\"u}ndogan, C.~Ams{\"u}ss, T.~C. Schmidt, and M.~W{\"a}hlisch.
\newblock {Content Object Security in the Internet of Things: Challenges,
  Prospects, and Emerging Solutions}.
\newblock {\em IEEE Transactions on Network and Service Management (TNSM)},
  19(1):538--553, March 2022.

\bibitem{hnkds-sunws-22}
R.~Hiesgen, M.~Nawrocki, A.~King, A.~Dainotti, T.~C. Schmidt, and
  M.~W{\"a}hlisch.
\newblock {Spoki: Unveiling a New Wave of Scanners through a Reactive Network
  Telescope}.
\newblock In {\em Proc. of 31st USENIX Security Symposium}, Berkeley, CA, USA,
  August 2022. USENIX Association.

\bibitem{intel-tinycrypt-20}
{Intel Corporation}.
\newblock {TinyCrypt Cryptographic Library}.
\newblock \url{https://github.com/intel/tinycrypt}, last accessed 07-17-2020,
  2017.

\bibitem{kblsw-pscli-21}
P.~Kietzmann, L.~Boeckmann, L.~Lanzieri, T.~C. Schmidt, and M.~W{\"a}hlisch.
\newblock {A Performance Study of Crypto-Hardware in the Low-end IoT}.
\newblock In {\em International Conference on Embedded Wireless Systems and
  Networks (EWSN'21)}, New York, USA, February 2021. ACM.

\bibitem{ksw-gpngi-21}
P.~Kietzmann, T.~C. Schmidt, and M.~W{\"a}hlisch.
\newblock {A Guideline on Pseudorandom Number Generation (PRNG) in the IoT}.
\newblock {\em ACM Comput. Surv.}, 54(6):112:1--112:38, July 2021.

\bibitem{libcose-21}
{Koen Bergzand}.
\newblock {libCOSE}.
\newblock \url{https://github.com/bergzand/libcose}, last accessed 05-10-2022,
  2018.

\bibitem{kscga-atcai-19}
D.~Kumar \emph{et~al.}.
\newblock {All Things Considered: An Analysis of IoT Devices on Home Networks}.
\newblock In {\em 28th {USENIX} Security Symposium ({USENIX} Security 19)},
  pages 1169--1185, Santa Clara, CA, Aug. 2019. {USENIX} Association.

\bibitem{ld-pedpm-19}
C.~Lachner and S.~Dustdar.
\newblock {A Performance Evaluation of Data Protection Mechanisms for Resource
  Constrained IoT Devices}.
\newblock In {\em International Conference on Fog Computing (ICFC '19)}, pages
  47--52, Piscataway, NJ, USA, 2019. IEEE.

\bibitem{RFC-2743}
J.~Linn.
\newblock {Generic Security Service Application Program Interface Version 2,
  Update 1}.
\newblock RFC 2743, IETF, January 2000.

\bibitem{mejlr-tslpi-20}
J.~Mades, G.~Ebelt, B.~Janjic, F.~Lauer, C.~C. Rheinl{\"a}nder, and N.~Wehn.
\newblock {TLS-Level Security for Low Power Industrial IoT Network
  Infrastructures}.
\newblock In {\em Design, Automation Test in Europe Conference Exhibition (DATE
  '20)}, pages 1720--1721, NJ, USA, 2020. IEEE.

\bibitem{mkw-huarca-18}
K.~Mindermann, P.~Keck, and S.~Wagner.
\newblock {How Usable Are Rust Cryptography APIs?}
\newblock In {\em International Conference on Software Quality, Reliability and
  Security (QRS '18)}, pages 143--.154, Los Alamitos, CA, USA, 2018. IEEE
  Computer Society.

\bibitem{stcdl-aruae-18}
P.~S. Munoz, N.~Tran, B.~Craig, B.~Dezfouli, and Y.~Liu.
\newblock {Analyzing the Resource Utilization of AES Encryption on IoT
  Devices}.
\newblock In {\em Asia-Pacific Signal and Information Processing Association
  Annual Summit and Conference (APSIPA ASC'18)}, pages 1200--1207,  2018. IEEE.

\bibitem{nzsr-pasei-22}
M.~Noseda, L.~Zimmerli, T.~Schläpfer, and A.~Rüst.
\newblock {Performance Analysis of Secure Elements for IoT}.
\newblock {\em IoT}, 3(1):1--28, 2022.

\bibitem{pkcs11-guide-14}
{OASIS Open}.
\newblock {PKCS \#11 Cryptographic Token Interface Usage Guide Version 2.40}.
\newblock
  \url{https://docs.oasis-open.org/pkcs11/pkcs11-ug/v2.40/cn02/pkcs11-ug-v2.40-cn02.html},
  last accessed 10-06-2021, 2014.

\bibitem{pkcs11-20}
{OASIS Open}.
\newblock {PKCS {\#}11 Cryptographic Token Interface Base Specification Version
  3.0}.
\newblock
  \url{https://docs.oasis-open.org/pkcs11/pkcs11-base/v3.0/os/pkcs11-base-v3.0-os.html},
  last accessed 10-06-2021, 2020.

\bibitem{openwsn-edhocc-22}
{OpenWSN}.
\newblock {EDHOC-C}.
\newblock \url{https://github.com/openwsn-berkeley/EDHOC-C}, last accessed
  05-10-2022, 2020.

\bibitem{phr-usadscl-19}
N.~Patnaik, J.~Hallett, and A.~Rashid.
\newblock {Usability Smells: An Analysis of Developers{\textquoteright}
  Struggle With Crypto Libraries}.
\newblock In {\em Fifteenth Symposium on Usable Privacy and Security ({SOUPS}
  2019)}, pages 245--257, Washington, D.C., Aug. 2019. {USENIX} Association.

\bibitem{plzdl-mhcis-19}
B.~Pearson, L.~Luo, Y.~Zhang, R.~Dey, Z.~Ling, M.~Bassiouni, and X.~Fu.
\newblock {On Misconception of Hardware and Cost in IoT Security and Privacy}.
\newblock In {\em 53rd International Conference on Communications (ICC '19)},
  pages 1--7, Piscataway, NJ, USA, 2019. IEEE.

\bibitem{sr-sidse-19}
T.~Schl{\"a}pfer and A.~R{\"u}st.
\newblock {Security on IoT Devices with Secure Elements}.
\newblock Technical report, WEKA, 2019.

\bibitem{linux-kconfig-20}
{The Linux Kernel Development Community}.
\newblock {Kconfig Language}.
\newblock
  \url{https://www.kernel.org/doc/html/latest/kbuild/kconfig-language.html},
  last accessed 28-09-2020, 2020.

\bibitem{um-wjdcwokc-18}
M.~Ukrop and V.~Matyas.
\newblock {Why Johnny the Developer Can't Work with Public Key Certificates: An
  Experimental Study of OpenSSL Usability}.
\newblock In {\em Topics in Cryptology -- CT-RSA 2018: The Cryptographers'
  Track at the RSA Conference 2018}, pages 45--64, Cham, 2018. Springer
  International Publishing.

\bibitem{wt-wjce-99}
A.~Whitten and J.~D. Tygar.
\newblock {Why Johnny Can't Encrypt: A Usability Evaluation of {PGP} 5.0}.
\newblock In {\em 8th {USENIX} Security Symposium ({USENIX} Security 99)},
  Washington, D.C., Aug. 1999. {USENIX} Association.

\bibitem{wolfssl-21}
{wolfSSL Inc.}
\newblock {wolfSSL Embedded TLS Library}.
\newblock \url{https://www.wolfssl.com/}, last accessed 10-20-2021, 2021.

\bibitem{RFC-2744}
J.~Wray.
\newblock {Generic Security Service API Version 2 : C-bindings}.
\newblock RFC 2744, IETF, January 2000.

\bibitem{zephyr-20}
{Zephyr Project}.
\newblock {Zephyr}.
\newblock \url{https://www.zephyrproject.org}, last accessed 07-17-2020, 2020.

\end{thebibliography}
\label{lastpage}

\end{document}